# Redefining Fuel Poverty: Introducing the Temporal Equity Framework (TEF)

**Torran Semple, John Harvey, Grazziela Figueredo, Lucelia Rodrigues, Mark Gillott, Phil Grunewald, Alexander Sullivan, Jan Rosenow**

## Abstract

The definition and quantification of fuel (or energy) poverty remain contentious, particularly within the English policy landscape. The current official definition—Low Income Low Energy Efficiency (LILEE)—systematically underestimates the condition in low- and middle-income households, effectively obfuscating changing energy affordability norms. This paper presents a sensitivity analysis of fuel poverty trajectories in Nottingham, UK, across competing definitions and diverging 'optimistic' (domestic efficiency upgrades) and 'pessimistic' (energy price inflation) scenarios. Results demonstrate that LILEE overstates the efficacy of efficiency upgrades while failing to adequately reflect the impact of price shocks. To address these deficiencies, we propose the Temporal Equity Framework (TEF), a budget standard-based approach paired with a proportional energy expenditure indicator that captures shifting affordability norms more equitably. The TEF also introduces continuous expenditure-based depth indicators to evaluate the entrenchment of fuel-poor households (the 'fuel poverty gap') and the resilience of non-fuel-poor households (the 'fuel poverty buffer'). We argue that responsive definitions which also recognise fuel poverty as a continuum of vulnerability, rather than a static, dichotomous state, are paramount to ensuring the equitability of energy transitions. Further, we suggest that the early identification of fuel poverty and spatial targeting of remedial resources could be markedly improved by adopting the TEF. We conclude that the adoption of a multidimensional budget standard-based definition is essential to facilitate a more accurate and just evaluation of energy transitions in England.

## 1. Introduction

Fuel (or energy) poverty occurs when a household lacks the financial means to fulfil its basic energy needs (Boardman, 1991). Its most acute symptom, prolonged exposure to cold indoor temperatures, remains a significant contributor to excess winter morbidity and mortality (Rudge & Gilchrist, 2007). Today, domestic energy provision is inextricably linked to the facilitation of a just energy transition: a political commitment to phasing out fossil fuel-based energies without exacerbating existing inequalities or precipitating new ones. While rhetoric around the fairness of energy transitions is occasionally limited to labour market displacement within the fossil fuel industry, the concept of a 'just transition' increasingly encompasses the broader societal impacts of decarbonisation and electrification policies, such as those addressed by the European Commission (n.d.) Social Climate Fund. In the context of fuel poverty, this means ensuring that the decarbonisation of the housing stock does not adversely impact the affordability of energy, especially among vulnerable populations (Abbasi, et al., 2022; Sovacool, et al., 2019).

The perceived 'fairness' of energy transitions, however, is contingent on underlying ideological tensions and the specific metrics used to evaluate social change. To interrogate this, we analyse the intertwined nature of fuel poverty quantification and energy transitions in England, focusing on the (in)compatibilities between current definitions and the measurement of equitable transitions. Specifically, we present a sensitivity analysis of fuel poverty trajectories according to various competing definitions in diverging economic scenarios. Of particular interest is the temporal responsiveness of existing measurement approaches, such as England's Low Income Low Energy Efficiency (LILEE) definition, to economic volatility and infrastructural transformation.

The LILEE definition was introduced as England's official fuel poverty definition in 2021 and considers a household to be fuel poor if its residual income falls below the poverty line after accounting for required energy costs, provided the home is deemed energy inefficient (see Table 1). *Sustainable Warmth* (BEIS, 2021), the policy paper underpinning this definitional transition, outlined a political agenda primarily focused on structural energy efficiency programmes. This included targeted retrofits for social and private housing, the expansion of the Energy Company Obligation (ECO) scheme, and enhanced home improvement grants (see BEIS (2021, p.7)). The specific stated objective of the LILEE definition was to 'better track progress toward the statutory fuel poverty target[1], whilst still reflecting the three key drivers of fuel poverty (low income, energy efficiency and prices)'. There is, however, growing concern that this framing disproportionately prioritises building characteristics over the immediate financial capacity of occupants (Semple, et al., 2024).

We argue that England's current LILEE definition is structurally ill-suited to transitional contexts characterised by energy price volatility. By design, LILEE's eligibility criteria are more concerned with energy efficiency than energy expenditure; consequently, the definition is largely insulated from macroeconomic shocks that drive energy costs upward. This structural insulation has significant governance implications: a definition that cannot register deteriorating affordability conditions cannot support effective monitoring, spatial targeting of remedial resources, or meaningful accountability for the distributional impacts of energy transitions. Our sensitivity analysis makes this concrete: in the 'pessimistic' scenario, mimicking the 2020–22 energy crisis, LILEE's incidence rises only marginally (an absolute change of 3.11 percentage points), while our proposed solution, the Temporal Equity Framework (TEF) is considerably more responsive.

The TEF is a dual-clause approach that leverages a tiered energy affordability indicator, inspired by the original 10% definition (see Boardman (1991) and Table 1), and a budget standard-based criterion, i.e. utilising Minimum Income Standard (MIS) budgets. The TEF also introduces novel continuous fuel poverty depth indicators, measuring entrenchment among fuel-poor households and resilience among non-fuel-poor populations. Findings indicate that the TEF approach is better positioned than existing definitions to improve the early identification of energy vulnerability, more effectively understand the dynamics of fuel-poor households, aid the spatial targeting of retrofits, and monitor transitional fairness.

[1] The fuel poverty target is to ensure that as many fuel-poor homes as is reasonably practicable achieve a minimum energy efficiency rating of Fuel Poverty Energy Efficiency Rating (FPEER) Band C, by 2030 (BEIS, 2021).

The paper has two primary aims: i) to quantify the divergence between competing definitions; and ii) to evaluate their responsiveness to bifurcating 'optimistic' and 'pessimistic' economic scenarios. To achieve this, we utilise the Synthetic Nottingham Homes (SNH) dataset (n=105,570) (Semple, et al., 2026), which provides household-level income, energy costs, and efficiency variables (see Section 4.1). The sensitivity analysis compares a range of existing definitions, which are introduced in the following section, against the novel TEF approach.

The remainder of this article is structured as follows: Section 2 presents necessary background regarding the English and wider UK policy contexts, examining existing fuel poverty definitions (Section 2.1), summarises the literature on the quantification of fuel poverty (Section 2.2) and explores how fuel poverty measurement approaches dictate the evaluability of equitable energy transitions (Section 2.3). Section 3 defines the TEF approach; Section 4 details the Synthetic Nottingham Homes (SNH) data and sensitivity analysis scenario parameters; Section 5 presents the sensitivity analysis results; and Section 6 offers policy recommendations, with a view to reconciling fuel poverty quantifications and the evaluation of equitable energy transitions in England and internationally.

## 2. Background

### 2.1. English fuel poverty politics

Modern fuel poverty discourse in England has been characterised by dissension and methodological uncertainty. The way that fuel poverty is quantified is a particularly polarising issue, leading to regular changes in the official definition. The initial 10% definition (Boardman, 1991; DEFRA, 2001) was succeeded by the Low Income High Costs (LIHC) indicator (Hills, 2012) in 2015, before LIHC was supplanted by the current LILEE definition in 2021 (BEIS, 2021).

It is noteworthy that this turbulent evolution of fuel poverty definitions is largely confined to England. Most comparable nations, including those across the UK and the Republic of Ireland, still use variants of the 10% definition (Irish Government, 2016; National Assembly for Wales, 2020; Northern Ireland Executive, n.d.). Scotland, however, is a notable outlier. The *Fuel Poverty Act 2019* (Scottish Parliament, 2019) introduced a dual-clause definition that considers a household to be fuel poor if energy costs exceed 10% of net household income *and* its residual income (after housing and energy costs) cannot facilitate an acceptable standard of living, based on the Minimum Income Standard (MIS). Enacted only two years prior to England's transition to LILEE, this legislation crystallised a widening divergence in the fuel poverty politics of Scotland and England. Table 1 provides the verbatim definitions for the 10%, LIHC, LILEE and Scottish 10% & MIS-based approaches.

*Table 1. Official fuel poverty definitions in the UK*

| Name | Description |
|---|---|
| 10% definition<br><br>(Boardman, 1991; DEFRA, 2001) | A fuel poor household is one which needs to spend more than 10% of its income on all fuel use and to heat its home to an adequate standard of warmth. This is generally defined as 21ºC in the living room and 18ºC in the other occupied rooms - the temperatures recommended by the World Health Organisation (WHO). |
| Low Income High Costs (LIHC) indicator<br><br>(Hills, 2012) | Households should be counted as fuel poor if a) they [have] required fuel costs above the median level; and b) were they to spend that amount they would be left with a residual income below the official poverty line. |
| 10% & MIS-based Scottish definition<br><br>(Scottish Parliament, 2019) | A fuel poor household is one where:<br><br>A) more than 10% (20% for extreme fuel poverty) of net income is required to pay for their reasonable fuel needs after housing costs have been deducted<br><br>**AND**<br><br>B) the remaining household income is not enough to maintain an acceptable standard of living, defined as at least 90% of the UK Minimum Income Standard (MIS) once childcare costs and disability or care benefits are deducted. |
| Low Income Low Energy Efficiency (LILEE) definition<br><br>(BEIS, 2021) | LILEE finds a household to be fuel poor if it has a residual income below the poverty line (after accounting for required fuel costs) and the home has an energy efficiency rating below [Fuel Poverty Energy Efficiency Rating—FPEER] Band C. |

Table 1 demonstrates the conceptual shift in English policy from an absolute measure (the fixed 10% threshold) to relative indicators (the median-tethered thresholds used by LIHC and LILEE). The political motivations underlying this period of instability are covered in detail elsewhere (Middlemiss, 2017; Semple, et al., 2025); hence, they are not discussed further here. This article instead examines the compositional implications of the shift from absolute to relative measures for household fuel poverty eligibility.

## 2.2. Literature review

The taxonomy of fuel poverty quantification is broadly divided into expenditure-based metrics, subjective or consensual (self-reported) indicators, and direct physical measurements, such as monitoring indoor temperatures (Rademaekers, et al., 2016; Thomson, et al., 2017). While emerging data streams like smart meter and financial footprint data show long-term promise (Cong, et al., 2022), they remain in their infancy; consequently, official fuel poverty definitions continue to rely on expenditure-based methods (Romero, et al., 2018) and will continue to do so for the foreseeable future. Beyond the definitions listed in Table 1, the literature has developed alternative cost-burden indicators, such as the two-times median proportional energy expenditure (2M) metric (Barrella, et al., 2021), which is effectively a dynamic version of the 10% definition, and the Hidden Energy Poverty (HEP) index (Antepara, et al., 2020; Meyer, et al., 2018), which captures energy self-rationing, i.e., underconsumption. Further, Minimum Income Standard (MIS)-based fuel poverty serves as an expenditure-based definition that captures fluctuations in living and energy costs (Moore, 2012), making fuel poverty contingent on wider financial pressures (e.g. inflation on necessities).

This paper narrows its empirical focus to five expenditure-based definitions deemed most relevant to the English policy context: the original 10% definition, the relative (median-tethered) LIHC and LILEE definitions, the official Scottish 10% & MIS (which utilises conjunctive AND logic), and a proposed disjunctive 10% | MIS variant (which operates on inclusive OR logic, as defined in Section 3). Focus on this specific set of definitions is intended to cover the complete evolutionary trajectory of official UK fuel poverty definitions, spanning absolute (10%), relative (LIHC and LILEE), and hybrid budget standard-based (MIS) frameworks. Comparing these methodologies across consistent conditions allows us to isolate how different mathematical formulations alter fuel poverty eligibility. Although we do not explicitly model a separate HEP metric in this analysis, the OR logic of the 10% | MIS variant likely captures cases of hidden energy poverty by explicitly accounting for the most vulnerable low-income, low-costs households.

### 2.3. Fuel poverty quantifications and equitable energy transitions

Fuel poverty definitions and the equitability of energy transitions are intrinsically linked. Decarbonisation requires the rearrangement of policy instruments and physical assets, which will in turn impact the distribution of end-user energy costs. Consequently, modern fuel poverty definitions must reflect shifting energy affordability norms. The LILEE definition's inability to capture such shifts is increasingly apparent. DESNZ (2025a) estimates that approximately 2.7 million households (11.0%) in England are fuel poor. Strikingly, the LILEE rate declined from 22.1% in 2010 to 11.0% in 2024, remaining largely insulated from the energy crisis (~13% in 2021/22 and 11.4% in 2023). This is an especially counterintuitive trend, given that typical domestic energy costs almost doubled between 2020 and 2023 (DESNZ, 2023). The Committee on Fuel Poverty (2024) mirror this sentiment, recommending that the LILEE metric 'be reviewed as it no longer captures the full range of households facing unaffordable bills'. By overlooking the effects of the recent cost-of-living and energy crises on household budgets, LILEE is ill-suited for gauging the fairness of energy transitions, precisely because it cannot account for changing energy affordability norms. Conversely, the original 10% definition is arguably outmoded; the 10% threshold is based on median household energy expenditure data from the 1990s (Boardman, 2009, p.22) and is increasingly arbitrary in the modern context. Furthermore, the LIHC indicator's dynamic, median-tethered thresholds absorb temporal shifts in energy affordability, resulting in artificially stagnant year-on-year rates (BEIS, 2021; Middlemiss, 2017).

To summarise, it is arguable that no existing English fuel poverty definition (10%, LIHC and LILEE), in isolation, can effectively track changing energy affordability norms. To address this, Section 3 introduces the Temporal Equity Framework (TEF): a multidimensional, budget standard-based approach related to the Scottish definition. Through the incorporation of complementary continuous fuel poverty depth indicators, the TEF moves beyond dichotomous classifications ('fuel poor' versus 'not fuel poor') to better identify both entrenched and borderline cases.

## 3. The Temporal Equity Framework (TEF)

The TEF approach quantifies the rate and depth of fuel poverty over time (see Figure 1) and is conceptualised as two independent processes: Step 1 is the primary fuel poverty classification and Step 2 is a set of complementary fuel poverty depth indicators.

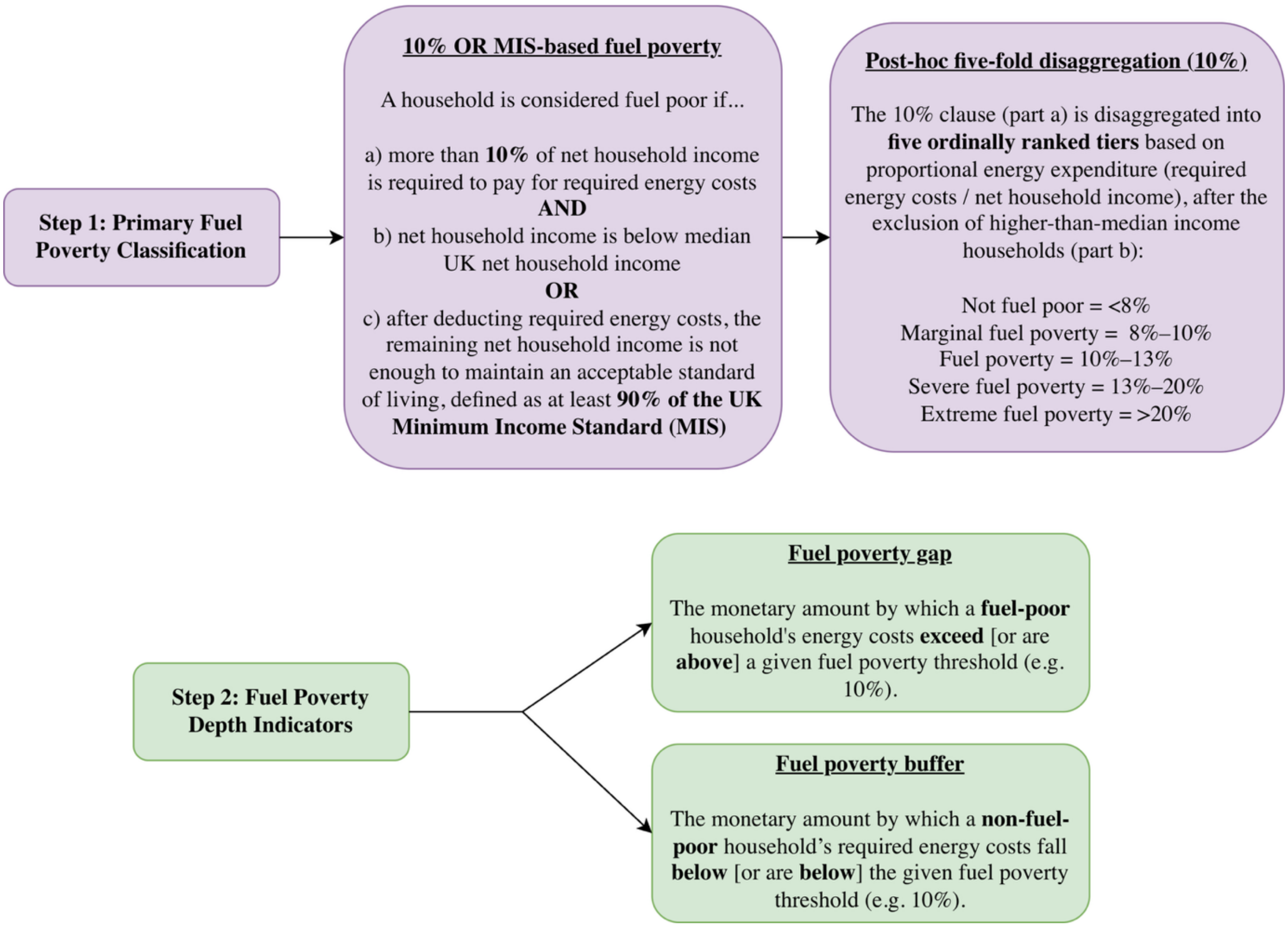

*Figure 1. The Temporal Equity Framework (TEF)*[2]

### 3.1. Step 1: Primary fuel poverty classification

Step 1 of the TEF integrates the traditional expenditure-based 10% definition with a budget standard-based clause. The framework utilises the Minimum Income Standard (MIS) (Loughborough University, Centre for Research in Social Policy, 2024) to evaluate the interaction between the increasing burden of energy costs and a household's capacity to maintain 'an acceptable standard of living' (Hirsch, et al., 2016; Scottish Government, 2017). This use of the MIS clause recognises that access to adequate domestic energy is increasingly a core tenet of living standards (Rao & Pachauri, 2017). Furthermore, this approach aligns with European Parliament (2023) guidance, which defines fuel poverty as 'a household's lack of access to essential energy services that provide basic levels and decent standards of living and health'. In the context of energy transitions, quantifying this relationship is vital for evaluating social adequacy. It should be noted that both the TEF and Scottish definitions utilise 90% of MIS, as opposed to the full (100%) MIS budget. The selection of a 90% threshold is grounded in empirical evidence and established policy convention. Evaluation of the relationship between MIS and lived poverty demonstrates that the risk of material deprivation accelerates non-linearly below 90% of MIS, establishing 90% as a robust statistical threshold for identifying genuine material hardship (Hirsch, et al., 2016; Scottish Government, 2017).

[2] Note: For the purposes of this analysis, 'net household income' refers to income after direct deductions (income and council tax) and housing costs (rent and mortgage payments). Further, for the calculation of MIS fuel poverty (Step 1: part c), MIS budgets exclude housing costs and council tax, as well as, childcare, water and fuel/energy costs, to match the Scottish Parliament (2019) MIS formulation (see Section 4.3: Table 4).

There are two notable distinctions between the TEF's primary fuel poverty classification and the Scottish definition (see Table 1). First, the TEF approach considers a household to be fuel poor if more than 10% of net household income is required to pay for energy (after housing costs) OR if the remaining income is insufficient to maintain an acceptable standard of living (below 90% of the MIS). Conversely, the Scottish definition operates on a stricter AND logic, requiring both thresholds to be met simultaneously. Our utilisation of the OR logic is specifically designed to capture instances of hidden fuel poverty (Meyer, et al., 2018). For example, under the Scottish framework, a household with required energy costs marginally below the 10% threshold that is simultaneously unable to maintain socially acceptable living standards would be excluded. We propose that energy costs remain an acute financial strain for households falling below the MIS threshold, irrespective of whether they meet the 10% clause. Hereafter, the Scottish AND formulation is referred to as '10% & MIS' and the TEF OR formulation as '10% | MIS'.

The second notable difference is the inclusion of a 'high-earner' cap in the TEF approach (see Step 1: part b in Figure 1). Specifically, households with a higher-than-median net household income are excluded from the primary classification if they only satisfy the 10% clause. However, in cases where this income cap is exceeded but the MIS threshold is breached, the household is still deemed fuel poor. The decision to include this cap reflects long-standing concerns that the 10% definition inadvertently captures high-income households with high running costs (Hills, 2012; Moore, 2012; Romero, et al., 2018) who are, in reality, financially secure enough to absorb these expenses. While the exact threshold for a 'high-income' cap is debatable, utilising median net household income provides a balanced benchmark. It is less restrictive than the standard relative poverty line (60% of median income), yet it successfully ensures that the classification process focuses predominantly on the bottom 50% of the income spectrum.

The TEF approach also includes a post-hoc disaggregation of the energy expenditure ratio (required energy costs over income) into five tiers of severity[3]. This formulation replaces an unnecessarily binary classification of fuel poverty with a spectrum of proportional energy affordability. It categorises households by energy expenditure ratios: 'not fuel poor' (<8%); 'marginal' (8–10%); 'fuel poor' (10–13%); 'severe' (13–20%); and 'extreme' (>20%), as previously considered by the Scottish Government (2017, p.59). This disaggregation incorporates themes of energy precarity (Petrova, 2018), identifying non-fuel-poor households on the threshold of poverty who are most vulnerable (among the non-fuel-poor population) to macroeconomic shocks. By recognising fuel poverty as a spatially and temporally dependent experience (Bridgen & Robinson, 2023; Thomson, et al., 2017), this framework allows for the identification of marginal cases, potentially reducing future remedial expenditure through pre-emptive intervention. Further, the distinction between fuel poverty (10–13%), severe (13–20%) and extreme (>20%) fuel poverty is potentially especially valuable for targeting scarce resources (Scottish Government, 2017) and tailoring interventions, whether delivered as standalone measures or combined support packages (e.g., direct payments, retrofits, subsidisation of low-carbon technologies) (Walker, et al., 2014). To elucidate the potential benefits of a tiered expenditure-based indicator over a binary counterpart, Table 2 sets out the policy utility, priority level, and illustrative interventions for each band

[3] As noted in Figure 1, the disaggregation applies to the capped version of the expenditure ratio, i.e. those with higher-than-median net household income are considered not fuel poor by default.

*Table 2. Five-fold 10% fuel poverty: interpretation, priority, political utility and policy response*

| Category | Interpretation & Priority | Political Utility & Policy Response |
|---|---|---|
| Not fuel poor (<8%) | Interpretation: Low risk of acute affordability distress.<br><br>Priority: Low. | Utility: Establishes a baseline for secure households; depth indicators (see Section 3.2) monitor intra-category dynamics.<br><br>Response: No direct intervention required. |
| Marginal (8–10%) | Interpretation: Precarious and highly vulnerable to price or income shocks.<br><br>Priority: Medium-Low (Preventative). | Utility: Identifies energy precarity before households cross statutory poverty lines.<br><br>Response: Pre-emptive measures (e.g., energy efficiency advice, low-cost insulation, smart tariffs). |
| Fuel poor (10–13%) | Interpretation: Moderate affordability strain; may be unidimensional vulnerability (e.g. precipitated by modest inefficiency or financial shortfall).<br><br>Priority: Medium. | Utility: Maintains longitudinal benchmark with legacy 10% statistics.<br><br>Response: Targeted standalone measures (e.g., seasonal energy support payments, minor retrofits). |
| Severe (13–20%) | Interpretation: Acute financial strain requiring substantial, multifaceted support.<br><br>Priority: High. | Utility: Captures deep, structural fuel poverty; enables targeted allocation of resources over lower-severity categories.<br><br>Response: Combined interventions (e.g., direct bill relief alongside retrofits). |
| Extreme (>20%) | Interpretation: Critical affordability strain; elevated risk of associated negative health consequences and likely compels spending trade-offs with other necessities.<br><br>Priority: Urgent. | Utility: Isolates the most vulnerable households; allows for prioritised allocation of potentially constrained resources to those most in need.<br><br>Response: Urgent, combined intervention: pairing emergency financial subsidies with deep-retrofit grants, coordinated across local housing and healthcare networks. |

## 3.2. Step 2: Fuel poverty depth indicators

Step 2 introduces depth indicators that quantify fuel poverty as a continuous monetary spectrum, enabling a granular assessment of both the depth of poverty (among the fuel-poor population) and household resilience (among the non-fuel-poor population). While Hills (2012) established the original 'fuel poverty gap', defined as the reduction in energy expenditure required to lift a household out of fuel poverty (based on LIHC poverty thresholds), current LIHC- and LILEE-based depth indicators are constrained by the structural limitations of their respective definitions (Semple, et al., 2024; DESNZ, 2025b). Critically, these conventional metrics ignore the non-fuel-poor population, failing to account for the resilience of these households to energy price volatility. In the TEF, household resilience is described as the 'buffer': the additional energy costs a household can absorb before crossing the 10% fuel poverty threshold. Figure 2 visualises the calculation of the 10% fuel poverty gap ($\Delta Y_G$) and buffer ($\Delta Y_B$) via the comparison of two hypothetical households ($H_1$ and $H_2$) with corresponding net household income (x-axis) and required energy costs (y-axis). The position of $H_1$ in the red-shaded segment (i.e. the 'gap zone') indicates that this household is 'fuel poor' according to the 10% definition; conversely, $H_2$, in the green-shaded area (the 'buffer zone'), is 'not fuel poor'. Because these depth indicators form a continuous spectrum of required costs over net household income, the higher-than-median income cap is omitted, as excluding households above the cap at this stage would artificially skew area-level depth statistics.

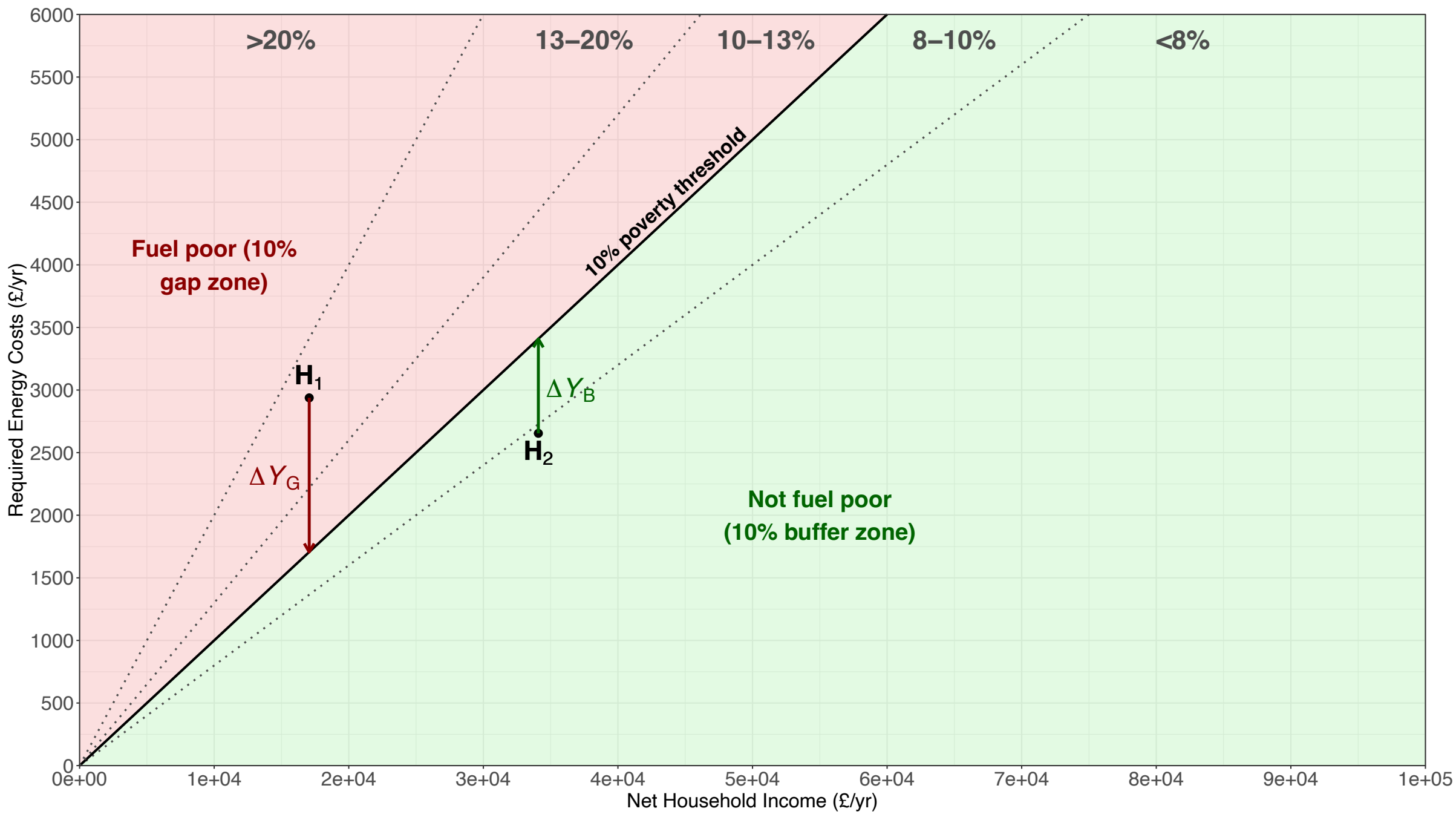

*Figure 2. The 10% fuel poverty gap and buffer[4]*

The monetary distance of $H_1$ and $H_2$ to the 10% poverty threshold is denoted by $\Delta Y$ and is calculated as follows:

$$\Delta Y = E_h - 0.1 I_h \tag{1}$$

Where $E_h$ is the required energy costs per household, h, and $I_h$ is net household income. For fuel-poor households, Equation 1 returns a positive value of $\Delta Y$ (i.e. the monetary amount required to lift a household out of poverty); conversely, not-fuel-poor households return a negative value (i.e. the monetary resilience to rising costs). Examining the distribution of $\Delta Y$ across Nottingham's Lower Super Output Areas (LSOAs)—small areas with approximately consistent population densities between 400 and 1,200 households and resident populations of 1,000–3,000 (ONS, 2024)—is a means of evaluating the spatiotemporal depth of fuel poverty in each of the sensitivity analysis scenarios (see Section 5.3).

[4] Note: Dotted lines in Figure 2 show the five-fold disaggregation of the 10% definition based on proportional energy expenditure. The 10% threshold is used for fuel poverty gap and buffer calculations in this analysis; however, the use of other expenditure thresholds (e.g. 8%, a potential proxy for energy precarity) may also be informative.

## 4. Methodology and data

This section details the Synthetic Nottingham Homes (SNH) dataset and the methods used for its generation. Subsequently, the rationale underlying the sensitivity analysis scenarios is outlined (Section 4.2). The sensitivity analysis was conducted in R (version 4.5.1); the SNH data, analysis code and environment dependencies are available in an anonymous OSF repository: https://bit.ly/OSF-TEF-Replication.

### 4.1. The Synthetic Nottingham Homes (SNH) dataset

Unit-level data (or microdata) pairing energy costs, household income and efficiency are often difficult to obtain due to privacy constraints. Consequently, various studies have synthesised microdata to overcome the unavailability of real-world information (Birkin & Clarke, 1989; Sakshaug & Raghunathan, 2014). We follow a similar approach, utilising a synthetic set of households that are broadly representative of the Nottingham population, containing paired energy costs, income and efficiency variables. While the SNH data are not of a particularly high fidelity, they still facilitate effective inter-methodological comparisons between competing fuel poverty definitions.

The SNH data were generated using an LSOA-specific direct sampling approach. For a given LSOA, $j$, let $Y_i$ represent the synthetic income for household, $i$. Let $\mu_j$ represent the mean LSOA-level income, based on ONS (2024) estimates, $\sigma_j$ the inferred LSOA-level standard deviation, based on the Understanding Society survey (University of Essex, Institute for Social and Economic Research, 2022), $U_j$ the estimated income upper bound and $N_j$ the number of households per LSOA. First, $N_j$ independent and identically distributed random values, $Z_i$, are generated from a standard normal distribution:

$$Z_i \sim N(0, 1) \; for \; i = 1, \dots, N_j \tag{2}$$

Given that household income often follows a log-normal distribution (Kloek & van Dijk, 1978), $Z_i$ is transformed into $Y_i$:

$$Y_i = exp\,(\mu'_j + \sigma'_j Z_i) \tag{3}$$

Where $\mu'_j$ and $\sigma'_j$ are expressed as:

$$\mu'_j = log\left(\frac{\mu_j^2}{\sqrt{(\sigma_j^2 + \mu_j^2)}}\right) \tag{4}$$

$$\sigma'_j = \sqrt{\left(log\left(1 + \frac{\sigma_j^2}{\mu_j^2}\right)\right)} \tag{5}$$

To ensure realistic income ranges, an upper bound is applied:

$$Y_i = min(Y_i, U_j) \; for \; i = 1, \dots, N_j \tag{6}$$

As LSOAs represent relatively homogenous small areas, unbounded log-normal sampling risks generating high-income outliers that are not representative of real-world conditions. To prevent this, an upper bound, $U_j$, is set dynamically for each LSOA:

$$U_j = \mu_j + 1.96\sigma_j \tag{7}$$

This bounding step truncates the synthetic income distributions at the 97.5th percentile. It is also worth noting that by capping income at 1.96 standard deviations above the mean, the model may understate the prevalence of genuine high-income households within economically heterogeneous LSOAs; however, such cases are of lesser importance to poverty classifications. The output of this process is LSOA-specific income microdata for Nottingham. The synthetic income data were then mapped to existing Energy Performance Certificate (EPC) records[5] (DLUHC, 2024), which contain dwelling characteristics (e.g. household energy efficiency and total floor area) needed to infer *required* energy costs. Required (or modelled) energy costs differ from actual energy expenditure and are a function of build characteristics and current energy unit price caps. While EPC data often overpredict actual consumption (Coyne & Denny, 2021; Few, et al., 2023), this approach ensures consistency across households by excluding occupant-driven behavioural variations, such as self-rationing. Specifically, required energy costs, $EC$, in GBP per year, are calculated as follows:

$$EC = GC + ElC \tag{8}$$

Where $GC$ and $ElC$ correspond to a household's required gas and electricity costs, respectively. The calculation of the $GC$ and $ElC$ components follow an equivalent process:

$$GC = \left(Q_{\text{total}} \cdot f_{\text{gas}} \cdot P_{\text{gas,cap}}\right) + 365 \cdot S_{\text{gas}} \tag{9}$$

Where $Q_{\text{total}}$ is a given household's total required consumption, an EPC-derived variable (DLUHC, 2024), $f_{\text{gas}}$ is the split factor, describing the typical split between gas and electricity consumption in a dual fuel household[6], $P_{\text{gas,cap}}$ is the unit price cap based on Ofgem (2025), with values of 6.34 pence per kWh for gas and 24.86 pence per kWh electricity. The daily standing charge, $S$, is multiplied by 365 to achieve an approximate annual figure: $S_{\text{gas}}$= 31.65 pence per day; $S_{\text{elec.}}$ = 60.97 pence per day (Ofgem, 2025).

## 4.2. Fuel poverty sensitivity analysis: baseline, Scenario 1 and Scenario 2 conditions

Sensitivity analysis involves the alteration of certain input parameters or assumptions to assess the effect on an output (Kunselman, 2023), and is especially common in the epidemiological (de Souza, et al., 2016; Schneeweiss, 2006) and economic literature (Dudel, 2017; Marques Junior, 2024). In this study, a sensitivity analysis is conducted to assess the impact of changing input parameters (e.g. household income, energy costs and energy efficiency) on fuel poverty rates, according to the 10%, LIHC, LILEE*, 10% & MIS (Scottish) and 10% | MIS (TEF) approaches.

---

[5] Note: See Semple et al. (2026) for further notes on the EPC mapping process.

[6] Note: Dual fuel was assumed in all cases. Based on Ofgem (2023), the typical ratio of gas to electricity consumption is 4.26; hence, gas typically accounts for ~80.99% of total consumption ($f_{\text{gas}} = 0.8099$) and electricity for ~19.01% ($f_{\text{elec.}} = 0.1901$).

Latterly, Step 2 of the TEF (continuous depth indicators) is also evaluated. The sensitivity analysis considers two distinct scenarios, as follows:

> **Scenario 1 (the 'optimistic' case):** Assumes all dwellings reach their 'potential energy efficiency' via comprehensive retrofitting (DLUHC, 2024), while energy unit prices and income remain fixed.
>
> **Scenario 2 (the 'pessimistic' case):** Mimics the 2020–2022 energy crisis, assuming energy costs outgrow income (+78.38% and +1.46% growth, respectively), while the energy efficiency of the housing stock remains fixed.

The conditions stipulated in these scenarios capture two relevant sets of diverging circumstances: an 'optimistic' scenario, characterised by significant upgrades to the housing stock, and a contrasting 'pessimistic' scenario, where energy expenditure (or fuel poverty) ratios would increase dramatically. The fuel poverty response across tested definitions is measured relative to a baseline scenario: the original conditions in the SNH data. Table 3 displays summary statistics for relevant parameters per scenario.

*Table 3. Sensitivity analysis: summary statistics for relevant variables (n=105,570)*

| Conditions | Median required energy costs | Median net household income | Median energy efficiency (SAP) | Median fuel poverty ratio (costs/income) |
|---|---|---|---|---|
| Baseline | £2,185 | £26,414 | 66 | 0.083 |
| Scenario 1 'optimistic' | £1,330 (↓) | £26,414 (→) | 81 (↑) | 0.050 (↓) |
| Scenario 2 'pessimistic' | £3,898 (↑) | £26,800 (↑) | 66 (→) | 0.145 (↑) |

Scenario 1 assumes that all dwellings in the SNH data achieve their 'potential energy efficiency', an EPC-derived variable representing the rating achievable if all plausible household-specific retrofits were implemented (DLUHC, 2024). The improved energy efficiency, specifically, the Standard Assessment Procedure (SAP) score, of the housing stock in this scenario requires the re-estimation of required energy costs. This is achieved by combining the household-specific potential energy consumption (DLUHC, 2024) with Ofgem (2025) energy unit price caps and daily standing charges, as previously described in Equations 8–9. Consequently, Scenario 1 conditions exhibit a considerable decrease in median modelled energy costs (£1,330/year) relative to the baseline (£2,185), with a concomitant reduction in the fuel poverty ratio.

Scenario 2 considers the effects of an increasing fuel poverty ratio. Since sophisticated energy price forecasting is inherently complex and often intractable, we assume a universal increase (+78.38%) in modelled energy costs across the sample. This rise mimics the actual proportional rise in median energy costs observed during the energy crisis: typical costs rose from £1,286 in 2020 to £2,294 in 2022 (DESNZ, 2023). Modest income growth (+1.46%) is also assumed to reflect the approximate increase in typical household income between 2020 and 2022 (ONS, 2023). A universal rise in energy costs would likely trigger energy self-rationing (Barrella, et al., 2022), as also observed in the UK during the cost-of-living and energy crises (Zapata-Webborn, et al., 2024). It is, therefore, noteworthy that behavioural responses to such conditions would likely prevent actual expenditure from reaching the theoretical (required) figures presented in this analysis.

### 4.3. Fuel poverty definition parameters: 10%, LIHC, LILEE*, Scottish (10% & MIS) and TEF (10% | MIS)

10% fuel poverty is simply the ratio of required energy costs to net household income: required costs in excess of 10% of income result in fuel poverty. LIHC fuel poverty is the cooccurrence of higher than median required energy costs and a residual household income below the poverty line (or threshold). LIHC estimates in this analysis utilise the 2021 poverty line, ~£18,840/year (ONS, 2022a), and a median energy costs threshold of £2,252/year (DESNZ, 2024). These thresholds were selected to match the input data (SNH variables) as best as possible: the SNH income variable is based on 2021 data (ONS, 2024), while EPC records are correct as of 2024 (DLUHC, 2024). Relatedly, LILEE fuel poverty is the overlap of household energy inefficiency, denoted by a Fuel Poverty Energy Efficiency Rating (FPEER) below 69, and residual income below the poverty line after required energy costs. FPEER is equivalent to a household's SAP score minus energy support payments (e.g. Warm Homes Discount). In this case, information regarding energy rebates, which is required to accurately calculate FPEER, was not available; hence, LILEE estimations are based on SAP score alone. As such, all LILEE estimates are referred to as 'LILEE*' (based on SAP rather than FPEER) from hereon. This omission is mitigated by DECC (2014) evidence that SAP and FPEER bands align in most instances, ensuring LILEE* serves as a robust proxy for LILEE.

The TEF, akin to the Scottish 10% & MIS-based approach, requires household composition information to apply Minimum Income Standard budgets (see Table 4; see also Appendix: Table 1 for exhaustive list of MIS budget components). As household composition is absent in the SNH data, we probabilistically assign categories based on Nottingham census data (ONS, 2022b) and floor area, *A*. The probability of a household, *i*, being assigned a composition type, *t*, was weighted to ensure a logical correlation between property size and occupancy (see Appendix: Table 2 for household composition categories and corresponding occupancy ranks[7]). The household composition assignment for each synthetic home is calculated as follows:

$$P(H = t_i) = \frac{p_t(1 + m \cdot A_{norm} \cdot r_t)}{\sum_{j=1}^{6} p_j(1 + m \cdot A_{norm} \cdot r_j)} \tag{10}$$

Where $p_t$ is a vector of Nottingham composition proportions, $A_{norm}$ is the min-max normalised total floor area, $m$ represents a constant multiplier (m=2), and $r_t \in \{1, 2, 3\}$ represents the occupancy rank assigned to the composition category, *t*. The heuristic selection of m=2 balances floor area–occupancy sorting against real-world census data (ONS, 2022b). While m=0 yields unweighted assignment and higher values (e.g. m=3) approach near-deterministic sorting, m=2 provides a maximum 7-fold weighting factor (1 + 2 · 1 · 3) for high-occupancy households in the largest properties (note: see Appendix: Table 2 for composition categories and corresponding occupancy ranks), compared to a 3-fold weighting factor (1 + 2 · 1 · 1) for single occupants. This suppresses improbable pairings (e.g. large families in small flats) while preserving more realistic rates of under-occupancy in larger properties.

[7] The household composition categories used by the ONS (2022b) and the MIS methodology (Loughborough University, Centre for Research in Social Policy, 2024) differ slightly; therefore, several of the ONS categories were aggregated to represent a single MIS category.

*Table 4. 2021 AHC MIS budgets (excl. childcare, rent, council tax, water and fuel costs) for different household compositions (Loughborough University, Centre for Research in Social Policy, 2024)*[8]

| MIS household composition categories | 2021 MIS budget (£/week) | 2021 MIS budget (£/year) |
|---|---|---|
| *Single* | 193.05 | 10,039 |
| *Couple* | 333.32 | 17,333 |
| *Single pensioner* | 173.91 | 9,043 |
| *Pensioner couple* | 275.69 | 14,336 |
| *Lone parent: 1 child* | 282.19 | 14,674 |
| *Lone parent: 2 children* | 359.98 | 18,719 |
| *Lone parent: 3 children* | 467.15 | 24,292 |
| *Couple: 1 child* | 375.82 | 19,543 |
| *Couple: 2 children* | 452.16 | 23,512 |
| *Couple: 3 children* | 538.48 | 30,341 |
| *Couple: 4 children* | 642.54 | 33,412 |

The calculation of TEF (10% | MIS) fuel poverty also requires median UK net household income for the higher-than-median earner cap (see clause b in Figure 1). For this we assume a value of £31,400, corresponding to median net household income in 2021 (ONS, 2022a).

## 5. Results and discussion

The sensitivity analysis results are structured as follows: Section 5.2 provides a unit-level overview of fuel poverty trajectories, detailing the frequency and proportion of affected households in each scenario, while Section 5.3 examines the dynamics of area-level (LSOA) fuel poverty. Baseline results (Section 5.1) are established first to contextualise the subsequent comparative analyses.

### 5.1. Baseline fuel poverty

Baseline conditions reflect the initial household income, energy costs and energy efficiency characteristics of the SNH data (see Table 3). Table 5 presents baseline frequencies and rates of 10%, LIHC, LILEE, Scottish (10% & MIS) and TEF (10% | MIS-based) fuel poverty. While baseline poverty rates appear high compared to those reported in national cohorts (Romero, et al., 2018), this reflects localised deprivation specific to the Nottingham-based sample. Nottingham is among the most deprived 6% of English districts (Nottingham City Council, 2025), meaning an elevated baseline is expected. Given that these rates would contract if recalculated at a national level, the value of this this analysis lies in the relative divergence between the definitions and the relative responsiveness to each set of conditions.

[8] Note: MIS budgets for 2021 are used for consistency with the SNH income data (based on 2021 data).

*Table 5. Baseline rates of 10%, LIHC, LILEE, Scottish (10% & MIS) and TEF (10% | MIS) fuel poverty (n=105,570)* [9]

| Definition | Fuel-poor households (n) | Fuel-poor households (%) |
|---|---|---|
| 10% | 38,883 | 36.83% |
| LIHC | 10,827 | 10.26% |
| LILEE* | 19,318 | 18.30% |
| Scottish 10% & MIS | 15,291 | 14.48% |
| TEF 10% \| MIS—capped | 38,131 | 36.12% |
| TEF 10% \| MIS—no cap | 41,672 | 39.47% |

The baseline fuel poverty rates diverged significantly depending on the definition employed, underscoring that different definitions measure fundamentally different conditions. Notably, the relative LIHC (10.26%) and LILEE* (18.30%) definitions yielded considerably lower rates of fuel poverty than 10%, but, perhaps surprisingly, a comparable rate to the Scottish 10% & MIS formulation. According to the capped 10% | MIS-based (TEF) definition, 36.12% (n=38,131) of households were classified as fuel poor, as opposed to 39.47% under the uncapped version; that is to say, 3.35% of the sample were omitted following the application of the high-earner cap.

The stark contrast between the conjunctive Scottish (10% & MIS) and disjunctive TEF (10% | MIS) formulations illustrates the key trade-offs for policy design. By requiring households to satisfy both the 10% and the MIS criteria, the Scottish approach is arguably too restrictive; in practice, it excludes a vast cohort of households classified as fuel poor under the 10% definition. This introduces a risk of obscuring disproportionately high energy costs among households that sit marginally above the MIS threshold. Conversely, the TEF formulation (10% | MIS) bypasses this limitation, capturing a broader spectrum of households meeting either the capped 10% threshold or the MIS criterion (after energy costs), thus ensuring the statistical eligibility of the most financially vulnerable populations in fuel poverty classifications.

As further illustration of the fundamental differences between absolute and relative (median-tethered) definitions, Figure 3 visualises discrepancies between the LILEE* and 10% definitions in baseline conditions. LILEE* fuel-poor households are those in the bottom left red-shaded quadrant; the remaining quadrants categorise households as Low Income High Energy Efficiency (LIHEE), High Income High Energy Efficiency (HIHEE) and High Income Low Energy Efficiency (HILEE) (BEIS, 2021). Each point (representing a household) is coloured according to its designation as 'fuel poor' or 'not fuel poor' according to the 10% definition. This secondary categorisation exposes the cohorts who are only selectively identified as fuel poor, depending on the definition applied.

[9] Note: In this case, TEF statistics only consider the 10% | MIS classification; the five-fold version of 10% fuel poverty is summarised for baseline, Scenario 1 and Scenario 2 conditions in Table 9. Step 2 of the TEF approach (depth indicators) is considered in Section 5.3.

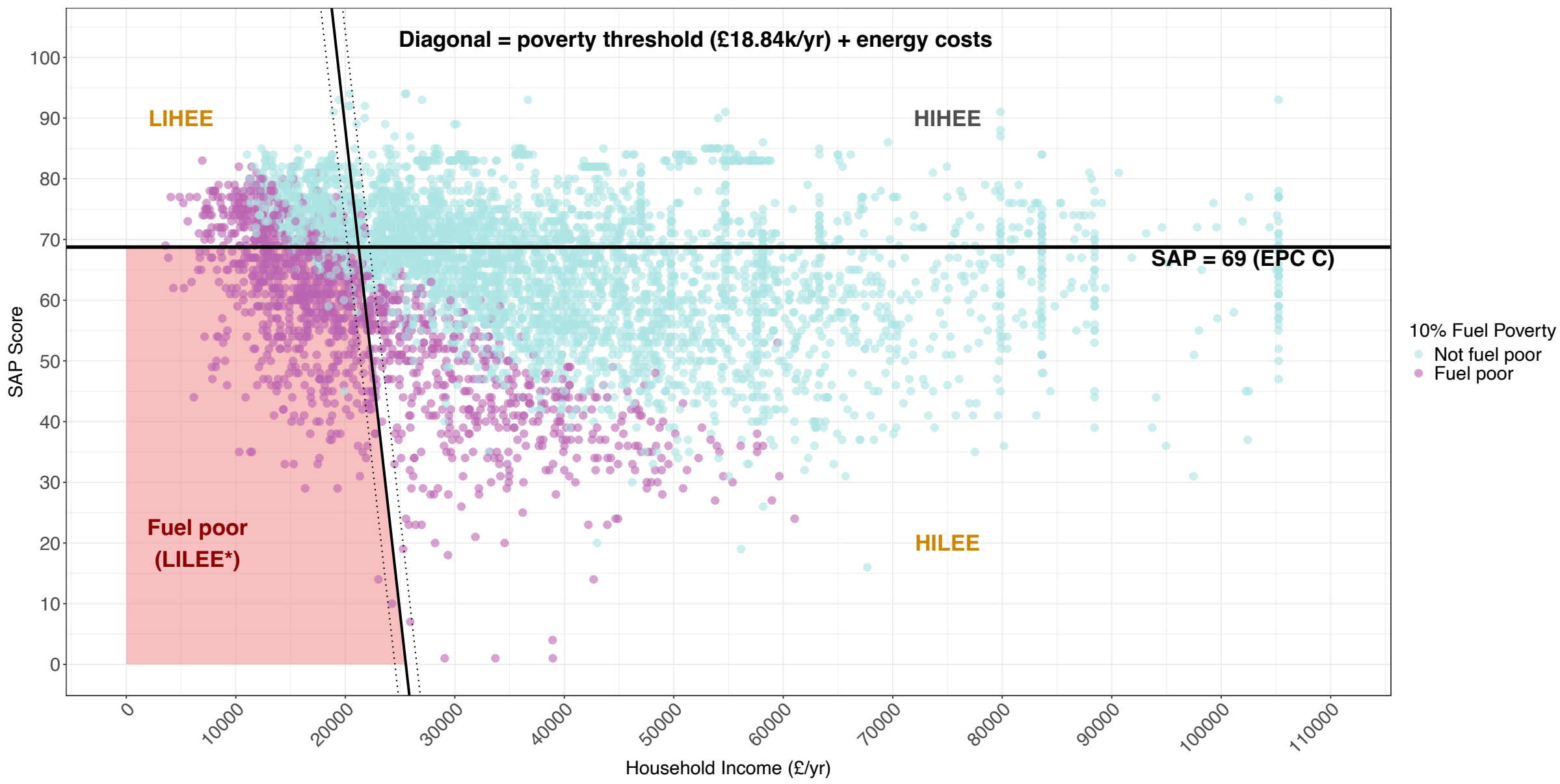

*Figure 3. Overlapping profiles of LILEE* and 10% fuel poverty in baseline conditions (n=5,313—10 randomly selected LSOAs in the SNH data)*

A more granular examination of the overlapping profiles of 10% and relative definitions (LIHC and LILEE) is presented in Table 6. The 10% fuel poverty rate is provided for each primary LILEE classification (i.e. those previously defined), and equivalent categories according to the nomenclature of LIHC (Hills, 2012): Low Income High Costs (LIHC), Low Income Low Costs (LILC), High Income Low Costs (HILC) and High Income High Costs (HIHC). This disaggregation is demonstrative of the types of households that are systematically overlooked by relative definitions. For example, a considerable proportion of LILC (63.39%) and LIHEE (52.39%) households (i.e. low-income households with relatively low running costs) are fuel poor according to the 10% definition but are ineligible according to LIHC and LILEE. The two rightmost columns present equivalent statistics for TEF (10% | MIS) fuel poverty. The rate of TEF fuel poverty is even more pronounced in LILC (74.02%) and LIHEE (67.22%) households, suggesting that the 10% definition may understate the financial burden of energy costs in such cases. In general, both sets of disaggregate statistics are demonstrative of the LIHC and LILEE methodologies' propensity to underestimate on-the-ground fuel poverty in low-income, energy-efficient households, a characteristic corroborated by previous studies (Middlemiss, 2017; Semple, et al., 2024).

*Table 6. Fuel poverty crosstabulations: LIHC and LILEE* vs. 10%; LIHC and LILEE* versus TEF (10% | MIS) (n=105,570)*

| **Quadrant** | **10% fuel poverty (%)** | | **TEF (10% \| MIS) fuel poverty (%)** | |
|---|---|---|---|---|
| | ***Not fuel poor*** | ***Fuel poor*** | ***Not fuel poor*** | ***Fuel poor*** |
| | **LIHC** | | **LIHC** | |
| LIHC (n=10,827) | 0.00% | 100.00% | 0.00% | 100.00% |
| LILC (n=26,251) | 36.61% | 63.39% | 25.98% | 74.02% |
| HILC (n=29,372) | 99.45% | 0.55% | 99.45% | 0.55% |
| HIHC (n=39,120) | 71.23% | 28.77% | 87.68% | 12.32% |
| | **LILEE*** | | **LILEE*** | |
| LILEE* (n=19,318) | 5.97% | 94.03% | 5.17% | 94.83% |
| LIHEE (n=17,760) | 47.61% | 52.39% | 32.78% | 67.22% |
| HIHEE (n=23,926) | 99.36% | 0.64% | 99.45% | 0.55% |
| HILEE (n=44,566) | 74.73% | 25.27% | 82.63% | 17.37% |

## 5.2. Sensitivity analysis results (I): unit-level results

This section evaluates the unit-level results of the sensitivity analysis. Table 7 displays competing rates of fuel poverty in Scenario 1 (representing the 'optimistic' case), where all households achieve their potential energy efficiency rating. Table 8 presents the results for Scenario 2 (the 'pessimistic' case), in which energy costs significantly outgrow income. In each table, the 'absolute change' column quantifies the percentage point difference between baseline and scenario conditions, while the 'relative change' column describes the responsiveness of each definition in relative terms.

### *Scenario 1 (the 'optimistic' case): unit-level results*

Scenario 1 was associated with a decrease in fuel poverty across all tested definitions. This is logical: higher household energy efficiency mitigates the financial burden of energy expenditure. Despite sustained criticism of efficiency-centric fuel poverty definitions (Middlemiss, 2017; Moore, 2012; Semple, et al., 2024), such as LIHC and LILEE, an extensive domestic retrofit programme is clearly a fundamental pillar of any long-term fuel poverty strategy. However, despite assumed upgrades, the 10% and TEF definitions still classify 16.19% and 22.10% of households as fuel poor, respectively. This reinforces a common critique of relative, energy-efficiency-focused definitions: the LIHC and LILEE approaches presume that energy (in)efficiency is the primary driver of and solution to fuel poverty, consequently omitting many low-income households with relatively low running costs. Conversely, the 10% and TEF approaches more robustly capture expenditure- and income-induced poverty, which is more directly linked to actual energy affordability. While the Scottish (10% & MIS) definition and TEF (10% | MIS) exhibit similar downward trends in relative terms, the stricter conjunctive logic of the former identifies a modest poverty rate, considering the socioeconomic standing of the study area (Nottingham City Council, 2025); however, this is merely a consequence of the low baseline rate. By contrast, the higher absolute prevalence maintained under the 10% | MIS formulation suggests that fabric upgrades alone cannot universally resolve the burden of energy costs; this formulation effectively exposes the residual income-induced fuel poverty that the Scottish framework masks.

*Table 7. Scenario 1 (the 'optimistic' case) results: 10%, LIHC, LILEE, Scottish (10% & MIS) and TEF (10% | MIS) fuel poverty (n=105,570)*

| Definition | Fuel poverty (%): Baseline | Fuel poverty (%): Scenario 1 | Absolute change | Relative change |
|---|---|---|---|---|
| 10% | 36.83% | 16.19% | –20.64% | –56.04% |
| LIHC | 10.26% | 2.51% | –7.75% | –75.54% |
| LILEE* | 18.30% | 3.74% | –14.56% | –79.56% |
| Scottish 10% & MIS | 14.48% | 9.27% | –5.21% | –35.98% |
| TEF 10% \| MIS—capped | 36.12% | 22.10% | –14.02% | –38.81% |
| TEF 10% \| MIS—no cap | 39.47% | 22.55% | –16.92% | –38.90% |

### *Scenario 2 (the 'pessimistic' case): unit-level results*

Scenario 2 results (Table 8) indicate that energy costs outgrowing income leads to increased fuel poverty across all tested definitions; however, the relative responsiveness of each definition varies substantially. The LILEE fuel poverty rate is least affected, increasing by only 3.11% (in absolute terms). This muted LILEE response in the 'pessimistic' case exemplifies the definition's failure to capture energy price volatility and the resultant financial burden on consumers. In other words, this is illustrative of LILEE's predisposition to identifying a low and inelastic rate of fuel poverty.

This characteristic implies that English fuel poverty rates are statistically destined to stagnate or decline, notwithstanding worsening economic conditions.

*Table 8. Scenario 2 ('the pessimistic' case) results: 10%, LIHC, LILEE, Scottish (10% & MIS) and TEF (10% | MIS) fuel poverty (n=105,570)*

| Definition | Fuel poverty (%): Baseline | Fuel poverty (%): Scenario 2 | Absolute change | Relative change |
|---|---|---|---|---|
| 10% | 36.83% | 78.51% | +41.68% | +113.17% |
| LIHC | 10.26% | 33.35% | +23.09% | +225.05% |
| LILEE* | 18.30% | 21.41% | +3.11% | +16.99% |
| Scottish 10% & MIS | 14.48% | 21.28% | +6.80% | +46.96% |
| TEF 10% \| MIS—capped | 36.12% | 57.34% | +21.22% | +58.75% |
| TEF 10% \| MIS—no cap | 39.47% | 78.67% | +39.20% | +99.32% |

A similar critique applies to the LIHC fuel poverty in Scenario 2. While the LIHC rate increased by over 200%, the absolute incidence (33.35%) may still be unduly conservative. Furthermore, the relative change in LIHC fuel poverty may be inflated by the assumption that the median energy costs threshold (£2,252/year) remains constant. Due to data constraints, calculating a dynamic UK-wide median was not feasible. If LIHC were re-estimated using scenario-adjusted median energy costs, the rate of LIHC fuel poverty would have been considerably lower and seemingly less responsive to 'pessimistic' conditions. Regardless, the 33.35% LIHC rate remains low in absolute terms compared to 10% (78.51%) and TEF (57.34%) fuel poverty.

The TEF (10% | MIS) produced a pronounced response in Scenario 2, identifying 57.34% of the sample as fuel poor. It is also noteworthy that the 10% and *uncapped* 10% | MIS definitions identified a similarly high proportion (~79%) of fuel-poor households. The gap between the capped and uncapped versions of the TEF demonstrates the operational utility of a high-earner cap. Specifically, 21.33% (~22,500 households) qualified as 10% fuel poor, but were excluded as their residual income exceeded the high-earner cap (median UK net household income), while also failing to satisfy the MIS criterion. Although the capped rate of 10% | MIS fuel poverty still appears high, the severity of the conditions proposed in the 'pessimistic' case warrants a commensurate response. In fact, the relative rise in 10% fuel poverty (+113.7%) is plausible, given that the national rate of 10% fuel poverty was estimated to have risen from ~18% in 2020 to ~37% in 2023 (DESNZ, 2024), a relative increase of ~106%.

A fuel poverty rate of ~79%, however, would imply that most low- and middle-income households were experiencing acute energy affordability challenges. While technically accurate, theories of targeted resource allocation (Hanna & Olken, 2018) demonstrate that overly broad poverty metrics are often structurally unfeasible, as expanding eligibility steeply reduces the level of support available per beneficiary. As such, we propose that the capped 10% | MIS definition is more likely to gain political traction, because it provides a better balance between the inclusive identification of fuel poverty and the realities of resource targeting. Conversely, the tendency of the LIHC and LILEE definitions to remain stable despite radical infrastructural and economic change may possess significant political expedience.

### ***Five-fold 10% fuel poverty***

A key aspect of the TEF approach is the post-hoc disaggregation of 10% fuel poverty into five discrete, ordinally ranked tiers, as shown in Table 9. This process produces higher-resolution information regarding the most severe and entrenched cases. Beyond the superior range of information produced, changing energy affordability norms are more accurately represented, and overreliance on the increasingly outmoded 10% threshold is reduced.

*Table 9. Five-fold 10% fuel poverty in baseline, Scenario 1 and Scenario 2 conditions (n=105,570)*[10]

| Conditions | Five-fold 10% fuel poverty | | | | |
|---|---|---|---|---|---|
| | **Not fuel poor: <8%** | **Marginal fuel poverty: 8–10%** | **Fuel poverty: 10–13%** | **Severe fuel poverty: 13–20%** | **Extreme fuel poverty: >20%** |
| Baseline | 54.44 | 12.08 | 13.76 | 13.98 | 5.74 |
| Scenario 1 | 75.65 | 8.61 | 7.55 | 6.11 | 2.08 |
| Scenario 2 | 41.21 | 2.34 | 7.77 | 22.52 | 26.17 |

The granularity offered by this spectral formulation is a marked improvement over its binary counterpart, providing enhanced insights into poverty severity and precarity (or resilience). For instance, the high concentration of severe (22.52%) and extreme (26.17%) cases in Scenario 2 reflects the acute financial burden imposed on end users. The binary 10% definition aggregates these groups, reductively categorising all cases as 'fuel poor'. Conversely, the five-fold version facilitates more accurate assessments of severity, offers greater utility for measuring temporal (in)stability, and captures subtleties of energy affordability among non-fuel-poor, yet precarious households (i.e. households classified as 'marginal' cases).

## 5.3. Sensitivity analysis results (II): area-level results

This section presents area (LSOA)-level results across all sets of conditions. The spatial response in LSOA-level fuel poverty in Scenario 1 and 2 conditions (versus the baseline) is represented graphically for each definition of interest: 10%, LIHC, LILEE, and the TEF (10% | MIS). The unit-level results are illustrative of the fact that the responsiveness of the Scottish (10% & MIS) and TEF (10% | MIS) formulations is comparable in relative terms (see Table 6 and Table 7). The fundamental divergence between these approaches lies in their absolute baseline rates rather than their sensitivity to changing conditions, therefore, replicating the area-level analysis for both would introduce redundant outputs. Consequently, the area-level evaluation focuses exclusively on TEF (10% | MIS) fuel poverty to ensure a more direct comparison with 10%, LIHC, and LILEE.

The provision of area-level results is essential to elucidate the spatial dynamics of fuel poverty, which are not revealed by the unit-level statistics alone. Figure 4 illustrates the interpretation of area-level shifts. The x-axis represents the baseline fuel poverty rate, while the y-axis is the fuel poverty rate in Scenario 1 or 2. The 'line of equality' serves as a reference point; LSOAs intersecting this line indicate zero response relative to the baseline. The position of points in the top-left (increasing fuel poverty) or bottom-right (reducing fuel poverty) segment reflects the extent of the fuel poverty response.

[10] Note: As previously defined in Figure 1, the five-fold disaggregation also includes the high-earner cap. Households exceeding this cap are assigned to the 'not fuel poor' column in Table 9 by default.

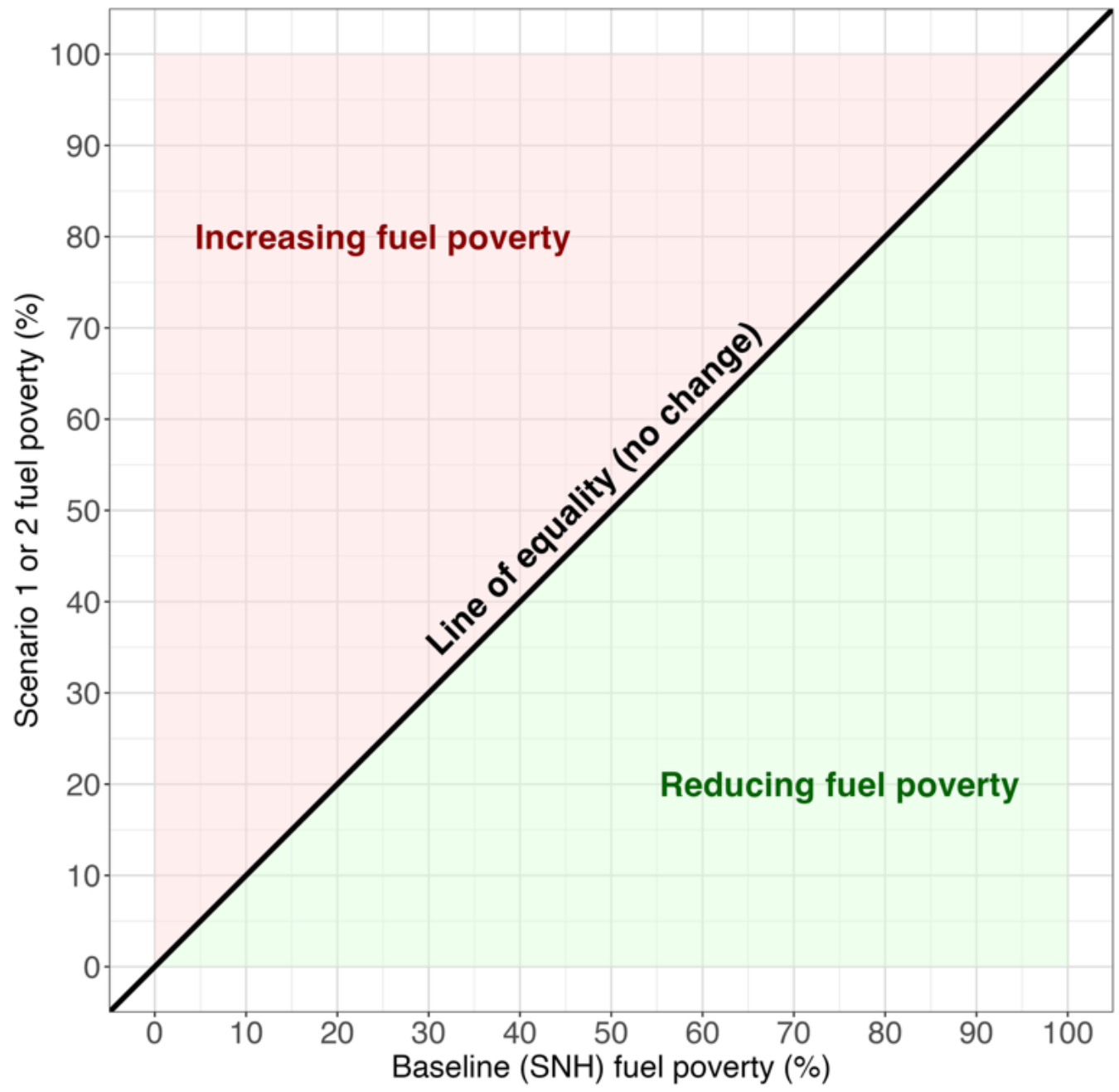

*Figure 4. The interpretation of LSOA-level fuel poverty shifts*

### *Scenario 1 (the 'optimistic' case): area-level results*

Scenario 1 resulted in a near-universal reduction in LSOA-level fuel poverty rates across all tested definitions (see Figure 5). Mirroring the unit-level results (see Section 5.2), this reduction is largely expected; however, the magnitude of the reduction fluctuates between the LSOAs, indicating that although energy efficiency programmes are a key remedial policy, their effectiveness can be spatially contingent. This spatial heterogeneity may be partially explained by 'retrofit poverty', where area-level characteristics, such as socioeconomic status and the proportion of privately rented dwellings, limit the feasibility of housing upgrades (Willand, et al., 2020). However, this analysis treats 'potential energy efficiency' as a purely building-based metric, independent of financial capacity; hence, the varying LSOA-level reductions are more likely attributable to the spatial distribution of older, 'hard-to-treat' properties (Mallaband, et al., 2013).

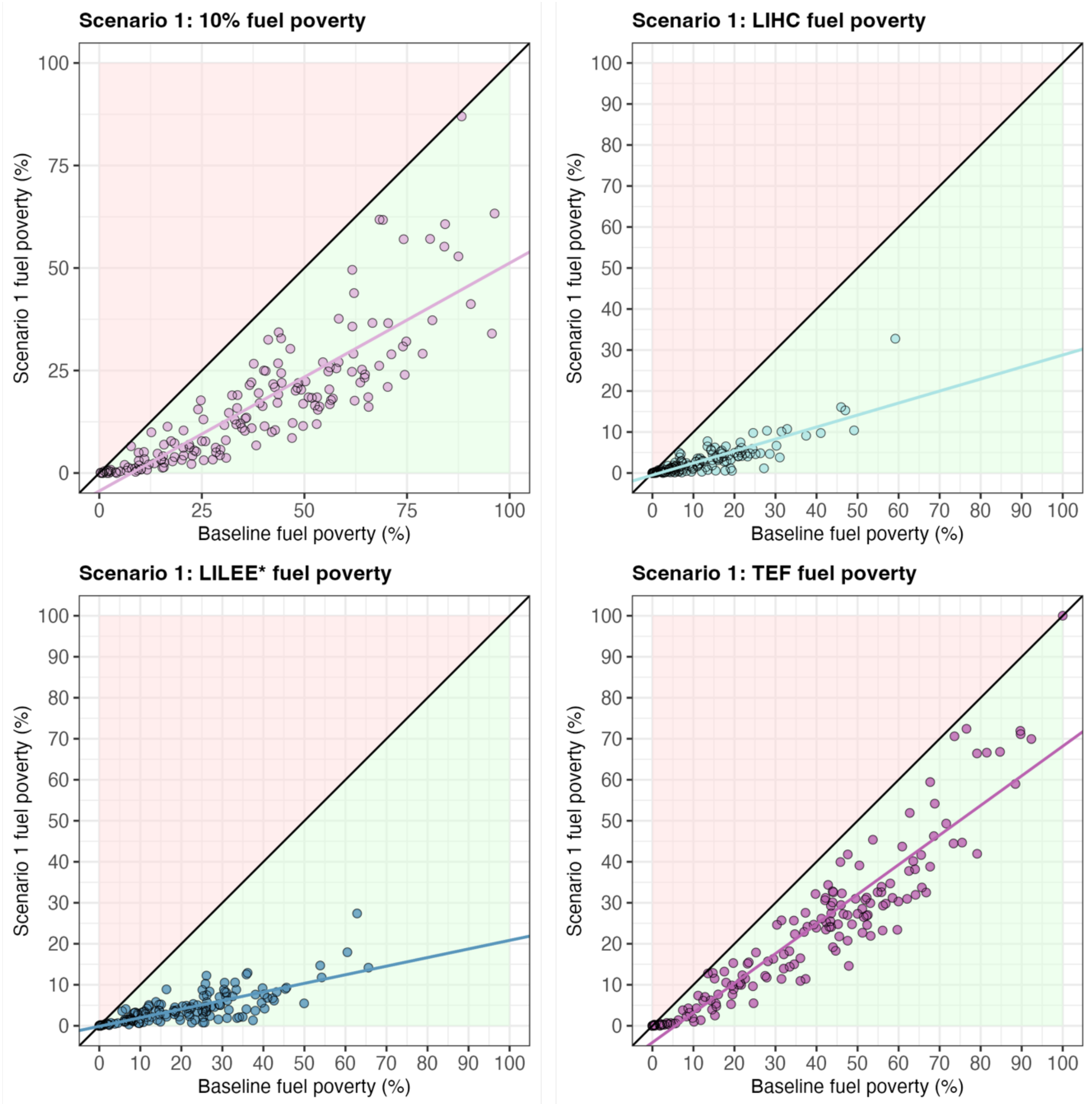

*Figure 5. LSOA-level fuel poverty in **Scenario 1** conditions (SNH data: n=164 LSOAs)[11]*

A clear dichotomy emerges between 10% and TEF (10% | MIS), and the relative definitions (LIHC and LILEE*). The line of best fit (the oblique, coloured line) in each panel is illustrative of the general area-level shift; for example, under the 10% definition, the line of best fit intersects the x-axis at approximately 10%, implying that LSOAs with baseline rates of 10% achieve near-total eradication (0%) in Scenario 1. The line of best fit is considerably more acute in the LIHC and LILEE plots, indicating a more pronounced reduction in fuel poverty. This further reinforces critiques of LIHC and LILEE as being primarily focused on energy efficiency as the main cause and solution to fuel poverty, while the actual affordability of energy is subordinate. Conversely, the 10% and TEF approaches generate similar trends: fuel poverty decreases sharply in some LSOAs but remains entrenched in others. This likely provides a more realistic representation of Scenario 1 conditions, since many LSOAs in the SNH data suffer high levels of income poverty (Nottingham City Council, 2025; ONS, 2024), which efficiency upgrades alone cannot overcome.

[11] Note: For comparability, axes were standardised (0–100), which leads to a clustering/overlap of LSOAs in the case of relative definitions (LIHC and LILEE*).

***Scenario 2 (the 'pessimistic' case): area-level results***

Scenario 2 simulates the 'pessimistic' case where energy costs outgrow household income at a rate of 78% (mimicking the 2021–2022 crisis). As anticipated, these conditions produced a pronounced rise in fuel poverty relative to the baseline.

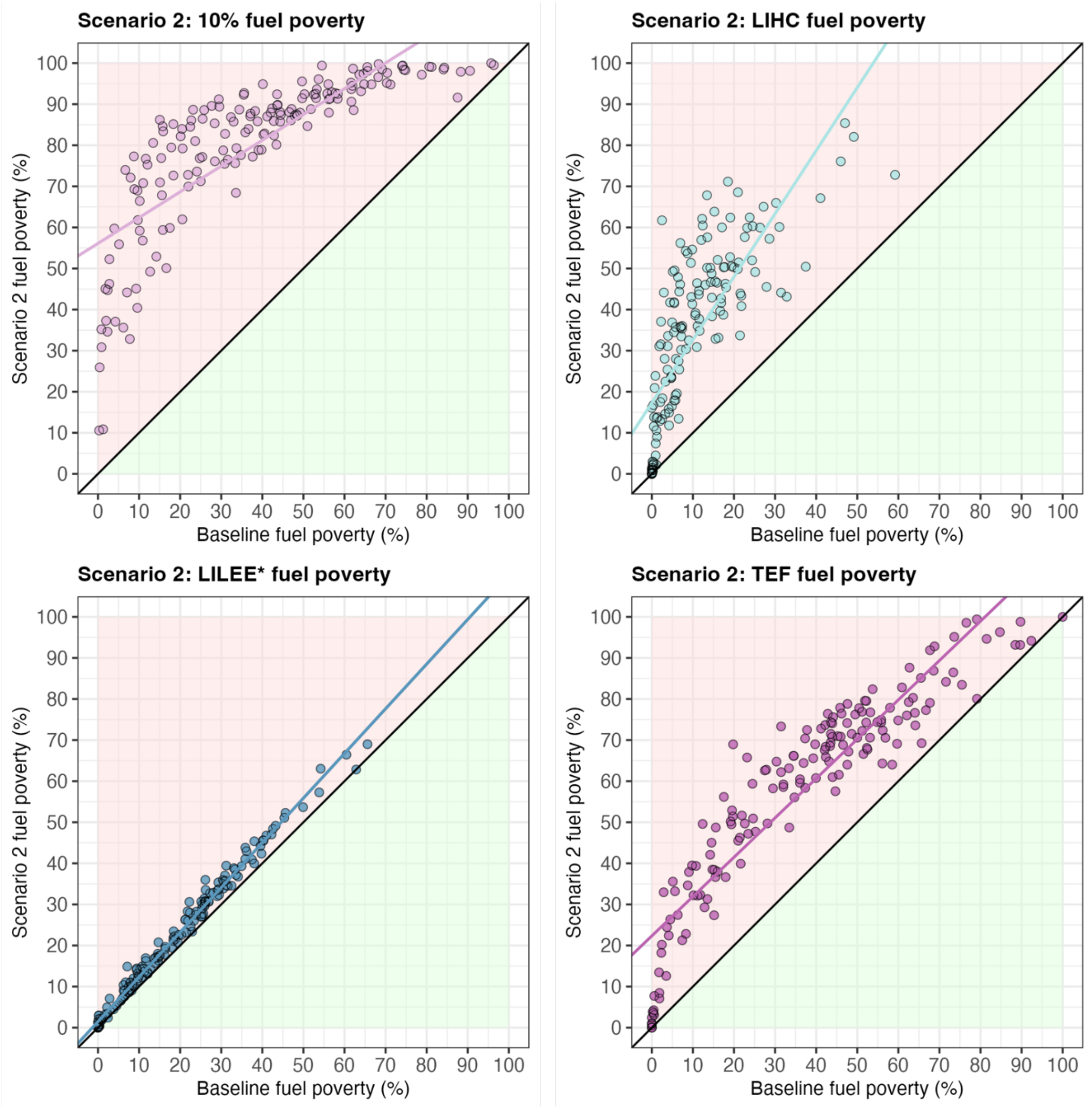

*Figure 6. LSOA-level fuel poverty in **Scenario 2** conditions (SNH data: n=164 LSOAs)*

Despite the universal increase in poverty, there are significant discrepancies between the definitions in terms of response magnitude. Notably, the current English definition (LILEE) produced the most anomalous response. In the Scenario 2 LILEE* plot, the line of best fit runs almost parallel to the line of equality, indicating an inelastic response despite severe economic conditions. In practice, this lack of responsiveness is concerning; the LILEE definition effectively obfuscates the seismic shift in energy affordability norms and its true impact on end users. In contrast, the 10% and TEF approaches produced substantial, elastic responses. While SAP was used as a proxy for FPEER (LILEE*), the equivalence of these metrics in most cases (DECC, 2014) suggests that this trend also reflects the response of England's official definition (BEIS, 2021;

DESNZ, 2026). Whether the responses of competing definitions are 'too great' or 'too little' remains a matter of ideological preference, as explored further in Section 6.

Although the sensitivity analysis is predicated on data for the English city of Nottingham, the findings maintain international relevance. Several European countries adopted median-tethered fuel poverty definitions following the introduction of LIHC (Palma, et al., 2024), while the European Energy Poverty Advisory Hub (EPAH) uses the 2M (two-times median proportional energy expenditure) indicator for cross-European comparisons. The EPAH, however, also note the multidimensionality of fuel poverty and the consequent need for a suite of absolute and relative indicators to capture its full extent (EPAH, 2020). Nevertheless, the limited responsiveness of median-tethered definitions to macroeconomic shocks and consequent insufficiency for measuring transitional fairness is likely to be internationally transferable.

***Depth indicators: the 10% fuel poverty gap and buffer***

Traditional fuel poverty definitions fail to capture a critical dimension of energy equality: the proximity of households to the poverty threshold. Table 10 presents summary statistics for the 10% fuel poverty gap and buffer (previously defined in Section 3.2) in each set of conditions. For brevity, statistics are provided for two contrasting LSOAs (n=1,007 households), while the 'Nottingham' row provides sample-wide (SNH) results. As discussed in Section 3.2, the 10% fuel poverty gap and buffer are based on the traditional 10% formulation; that is to say, the full uncapped spectrum of energy expenditure ratios is used to calculate associated area-level depth statistics.

*Table 10. 10% fuel poverty gap and buffer in example LSOAs and Nottingham*[12]

| **Area** | **Baseline** | | | **Scenario 1 'optimistic'** | | | **Scenario 2 'pessimistic'** | | |
|---|---|---|---|---|---|---|---|---|---|
| | **Fuel poor** | **Med. gap** | **Med. buffer** | **Fuel poor** | **Med. gap** | **Med. buffer** | **Fuel poor** | **Med. gap** | **Med. buffer** |
| **E01013965** | 65.6% | 709 | 499 | 16.1% | 591 | 1123 | 98.1% | 2628 | NA |
| **E01013977** | 0.08% | NA | 2721 | 0% | NA | 4181 | 35.2% | 972 | 1593 |
| **Nottingham** | 36.8% | 597 | 1081 | 16.2% | 396 | 1579 | 78.5% | 1607 | 946 |

LSOA 'E01013965' belongs to the lowest Index of Multiple Deprivation (IMD)[13] quintile, ranking it among the most deprived 20% of English LSOAs, whereas 'E01013977' is relatively affluent (DLUHC, 2020). The socioeconomic separation of these two LSOAs results in contrasting depth statistics. The 'deprived' LSOA showed significant improvement in Scenario 1, with a reduced gap and expanded buffer, indicating improved conditions for fuel-poor households and increased resilience among non-fuel-poor households, respectively. Conversely, the fuel poverty gap widened dramatically in Scenario 2, illustrating how macroeconomic shocks deepen poverty in already-deprived areas. The 'affluent' LSOA exhibits a distinct trajectory: the buffer was initially high and grew further in Scenario 1. However, the buffer contracted to £1,593 in Scenario 2, coinciding with a significantly increased fuel poverty rate (35.2%). This suggests that already-deprived LSOAs are disproportionately affected in the 'pessimistic' case because the existing depth of poverty and the precarity of non-fuel-poor households in these areas is more acute. Relatively affluent areas, by contrast, possess a substantial financial buffer, insulating them from economic shocks.

[12] Table 10 notes: all gap and buffer statistics are monetary values (i.e. GBP) rounded to the nearest integer; the 'fuel poor' column refers to the proportion of fuel-poor homes under the 10% definition, *without* the high-earner cap; 'NA' values in Table 10 indicate cases where the fuel poverty rate is too low (<2.5%) or too high (>97.5%) for robust median statistics.

[13] IMD is a composite deprivation indicator that is used to rank English LSOAs from least to most deprived (DLUHC, 2020).

Figure 7 visualises the $\Delta Y$ monetary depth spectrum (defined in Figure 2), illustrating the continuous distribution of household energy resilience (buffer) and depth (gap) across the two example LSOAs. The $\Delta Y$ distributions corroborate earlier findings that deprived areas are characterised not only by more entrenched and severe fuel poverty but also by a pervasive lack of resilience to volatile energy costs (i.e. heightened energy precarity). A failure to recognise that these areas possess lower resilience and a heightened sensitivity to macroeconomic shocks obscures the reality that energy price volatility yields deeply uneven distributional social impacts.

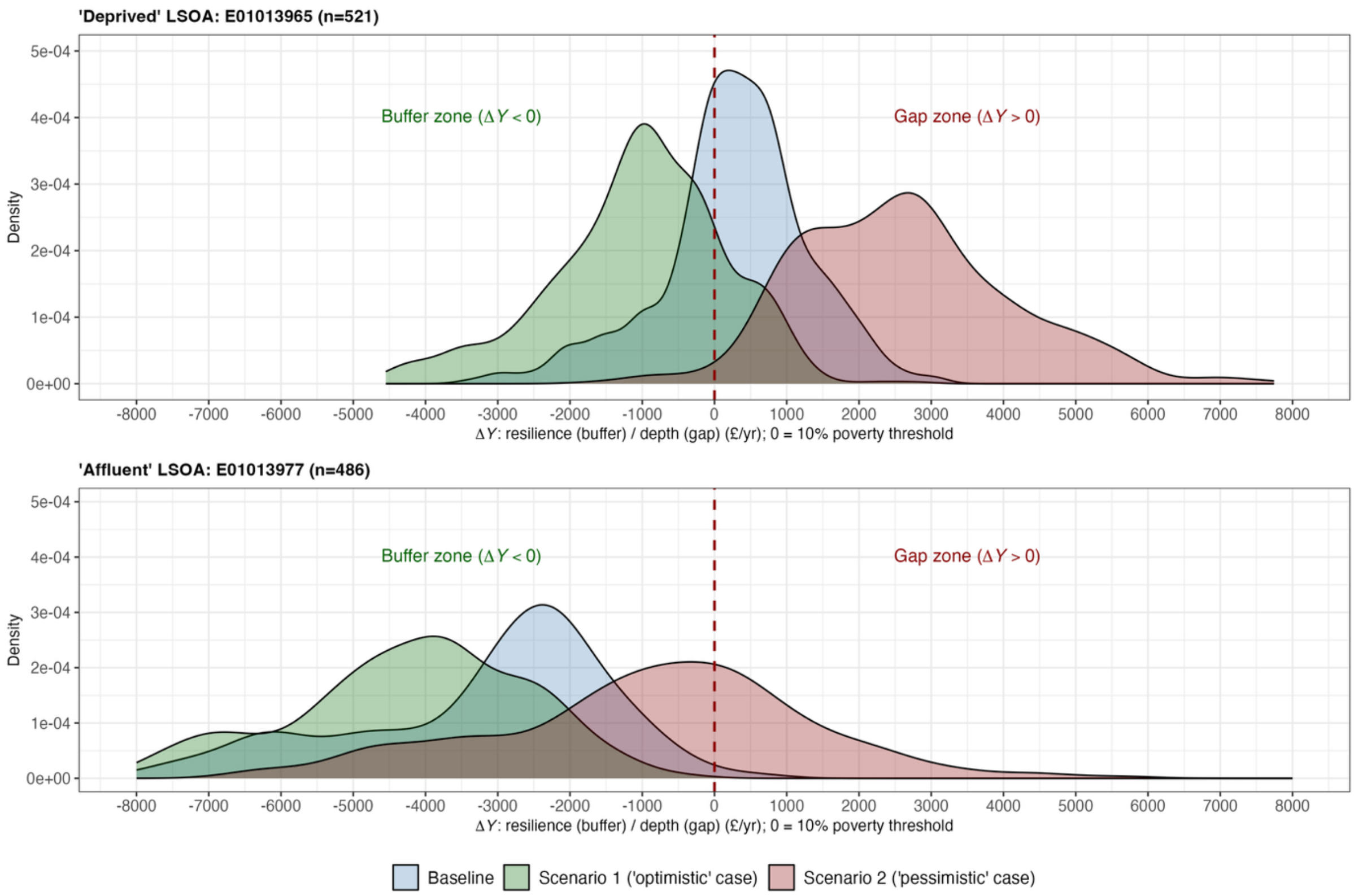

*Figure 7. Delta Y distributions in example 'deprived' and 'affluent' LSOAs*

## 6. Conclusions, policy implications and limitations

### 6.1. Conclusions

This study evaluated the performance of competing fuel poverty definitions in two distinct scenarios, representing 'optimistic' and 'pessimistic' sets of diverging conditions. The results provide compelling evidence that median-anchored fuel poverty definitions like LIHC and LILEE do not adequately reflect the full extent of fuel poverty or shifting energy affordability norms. Given that energy transitions will inevitably impact end-user energy costs, this paper argues that the LILEE definition is structurally ill-suited to transitional contexts characterised by price volatility. Interrogation of the 'pessimistic' scenario exposed a critical flaw in the LILEE definition: it statistically precludes fuel poverty from rising to any meaningful extent, notwithstanding fluctuations in energy prices or the cost of living more generally.

In response to the deficiencies of LILEE, we propose the Temporal Equity Framework (TEF) as a robust alternative for quantifying fuel poverty throughout energy transitions. The primary classification within the TEF approach is informed by the 10% | MIS (budget standard-based) definition, wherein households unable to achieve a 'socially acceptable standard of living' after energy expenses, are recognised as fuel poor. This stands in stark contrast to the LILEE definition, which systematically obscures the financial burden of energy costs among low- and middle-income households. While inspired by the Scottish 10% & MIS definition, the TEF approach instead utilises a disjunctive (10% | MIS) logic, which intends to capture cases of hidden fuel poverty, for example, cases in which required energy expenditure falls short of the 10% threshold, but residual income is not enough to satisfy the MIS criterion. Another of the TEF's notable deviations from the Scottish definition is that the 10% clause is disaggregated into five tiers of severity; this scalar approach identifies severe and entrenched cases, while additional monitoring of 'marginal' populations may assist policymakers in preventing precarious populations from becoming poor. By rejecting the traditionally binary classification ('fuel poor' versus 'not fuel poor'), the TEF offers a higher-resolution scale that better reflects lived experiences.

A further advantage of the TEF pertains to its relative responsiveness to external conditions, especially compared to the incumbent LILEE definition. The area-level results demonstrated the tendency of the LILEE definition to overstate the benefits of energy efficiency upgrades in the 'optimistic' case, while trivialising the increased financial burden of energy costs in the 'pessimistic' case. Conversely, the TEF approach produced what we consider to be more realistic outcomes. The reduction in area-level fuel poverty was more modest in the 'optimistic' case and varied spatially across LSOAs, recognising that energy efficiency upgrades are not a universal solution to energy unaffordability. In the 'pessimistic' scenario, the TEF produced a pronounced response, identifying approximately 57% of the SNH sample as fuel poor. Such a rate of fuel poverty is conceivable, considering the extreme economic circumstances proposed, and particularly given that Nottingham is among the most deprived municipalities in England. The perceived acceptability of any fuel poverty definition, however, is intrinsically laden with ideological implications. Ultimately, the political expedience of a poverty metric that yields a stagnant and relatively low rate of deprivation, such as LILEE, may possess more immediate utility for policymakers than elastic measures that fluctuate sharply in response to economic volatility. By decoupling the official fuel poverty definition from macroeconomic shocks, such metrics present a semblance of stability that may, in practice, obfuscate the impact of volatile costs on lived experiences.

The TEF approach also introduces refined fuel poverty depth indicators: the fuel poverty gap and buffer. While inspired by the Hills (2012) LIHC-based fuel poverty gap, the TEF indicators are anchored to Boardman's (1991) 10% poverty threshold, serving as critical temporal indicators of entrenchment, precarity and energy equality. As practically deployable metrics, these facilitate the measurement of geographical shifts in energy equality over time. This longitudinal and spatial lens is essential, as volatile energy prices, whether driven by decarbonisation or unforeseeable macroeconomic shocks, will undoubtedly continue to shape the lived experience of fuel poverty.

## 6.2. Policy implications

The central policy implication of this study is that the official fuel poverty definition in England requires urgent reconsideration. Specifically, we recommend that the TEF approach, utilising the five-fold 10% | MIS classification process in conjunction with complementary depth indicators, would provide a more representative assessment of fuel poverty in England. The utilisation of such a measurement framework is essential to rigorously evaluate the equity of energy transitions on the route to net zero. We demonstrate that LILEE is structurally incapable of representing the increasing financial burden that could result from a transition characterised by energy price volatility. This has significant governance implications for the immediate- and longer-term identification, monitoring and remediation of fuel poverty. Furthermore, the UK Government's accountability for gauging the potential harms and inequities induced by energy transitions is limited by the LILEE definition. The fact that three definitions (10%, LIHC and LILEE) have been employed in England since 2015 may render further methodological oscillations politically unattractive. However, the fairness of energy transitions must be measurable via robust metrics, one of which is the official fuel poverty definition. Consequently, the statistical rationale for an amended or new approach to measuring fuel poverty in England is increasingly irrefutable.

Beyond the official fuel poverty definition, the findings of the sensitivity analysis provide insight into how specific dimensions of fuel poverty might be remedied. Specifically, the divergent responses across LSOAs in the 'optimistic' case, where some experienced drastic decreases in fuel poverty and others remained stable, are illustrative of entrenched spatial inequalities, such as 'retrofit poverty' or 'hard-to-treat' homes. In cases where retrofit poverty or other structural inequalities affect the viability of housing upgrades, targeted income support must continue to complement the remediation of fuel poverty. The targeting of such support, as well as decisions over the type of support that is required, could be enhanced by the TEF approach. The TEF allows for earlier identification of areas with low resilience to energy price volatility, which could profitably be used to spatially target preventative strategies. Perhaps most crucially, the TEF also provides a better basis for measuring transitional fairness across the entire continuum of energy vulnerability, such that changing energy affordability norms are represented in terms of depth (among fuel-poor households) and resilience (among non-fuel-poor populations).

Although we primarily consider the English policy landscape, the findings have international relevance: median-tethered and efficiency-centric fuel poverty indicators are insufficient measures of transitional fairness. This is particularly applicable to European nations that still adopt median-tethered definitions in isolation, and highlights a potential limitation of the EPAH 2M approach. To summarise, relative fuel poverty definitions, especially when deployed in isolation, fail to capture the full extent of the condition and do not fairly represent the potential for energy costs to become an increasing financial burden during the transition. This characteristic is increasingly at odds with the EU's 2023 Energy Efficiency Directive, which urges multidimensional measurement solutions and the incorporation of living standards-based criteria. The TEF, in contrast, directly accounts for living standards and external cost-of-living pressures via the disjunctive 10% | MIS formulation, allowing fuel poverty eligibility to respond proportionately to economic volatility and radical infrastructural change.

### 6.3. Limitations and further work

Several limitations of this analysis warrant acknowledgement. The estimation of MIS-based fuel poverty necessitated the probabilistic assignment of household composition to existing SNH data, a process that is inherently reductive. Nonetheless, the advantages of the 10% | MIS-based formulation, as opposed to 10% in isolation, remain demonstrable. A further limitation concerns the reliance on required energy costs in lieu of actual energy expenditure data. As documented in previous studies, required energy costs may overestimate consumption, especially during periods of high energy and living costs. Consequently, the energy costs variables used in the sensitivity analysis, particularly in the 'pessimistic' scenario, likely overestimate actual expenditure, as behavioural responses (i.e. self-rationing) were not considered. The recalculation of the TEF's 10% | MIS-based classification using *actual* energy expenditure offers a valuable direction for further research. The static nature of the UK-wide median energy costs threshold, as utilised by the LIHC indicator, in the baseline, optimistic and pessimistic scenarios must also be noted. In actuality, this threshold would fluctuate in the tested scenarios, leading to a more stable LIHC rate (less deviation between scenarios); however, as the analysis is specific to Nottingham, it was not possible to infer a revised UK median costs threshold per scenario.

Further work should consider fuel poverty trajectories based on real-world forecasts, given that the scenarios presented here are intentionally dichotomous, largely theoretical, and specific to the city of Nottingham. The generalisability of findings is also constrained by the geographical focus and the stylised nature of the considered scenarios. As such, further work may also consider scaling up TEF fuel poverty estimates to regional or national level estimates. It may also be advantageous to explore the operational feasibility of implementing an MIS-based definition in England. This study primarily focuses on the conceptual and statistical rationale for an amended fuel poverty measurement framework; accordingly, further investigation into the associated data infrastructure is required to facilitate the potential integration of the TEF into official national statistics. Further empirical exploration of the 'high-earner' is also necessary. While the current iteration of the TEF relies on the median UK net household income to restrict eligibility to the bottom 50% of the income distribution (contingent on the MIS criterion also being satisfied), further research is required to provide robust statistical validation for this choice. Specifically, further analyses may attempt to evaluate whether median income acts as a meaningful inflection point.

**Acknowledgments**

The authors would like to thank Dr Brenda Boardman for her expert input and feedback on an initial draft of this article. This work was supported by the UK Engineering and Physical Sciences Research Council (EPSRC) grant EP/S023305/1 for the University of Nottingham, EPSRC Centre for Doctoral Training in Horizon: Creating Our Lives in Data, the Energy Demand Observatory and Laboratory (EDOL) grant EP/X00967X/1, and the Medical Research Council grant MR/Z505390/1.

## Appendix

*Appendix: Table 1. MIS budget components (£/week) per household composition (Loughborough University, Centre for Research in Social Policy, 2024)*

| **2021 MIS budget component categories** | *Single* | *Couple* | *Single pensioner* | *Pensioner couple* | *Lone parent: 1 child* | *Couple: 1 child* |
|---|---|---|---|---|---|---|
| Food | 50.99 | 83.13 | 47.10 | 75.25 | 57.91 | 89.90 |
| Alcohol | 6.10 | 12.20 | 8.05 | 12.37 | 5.40 | 10.47 |
| Tobacco | 0.00 | 0.00 | 0.00 | 0.00 | 0.00 | 0.00 |
| Clothing | 9.46 | 18.93 | 7.91 | 15.82 | 23.43 | 33.64 |
| Water rates | 6.09 | 6.09 | 6.09 | 7.18 | 10.37 | 10.37 |
| Council tax | 18.68 | 24.93 | 18.68 | 24.93 | 21.82 | 29.08 |
| Household insurances | 1.57 | 1.76 | 1.48 | 1.55 | 1.27 | 1.32 |
| Fuel | 13.50 | 16.73 | 11.01 | 15.07 | 15.93 | 17.27 |
| Other housing costs | 1.48 | 1.48 | 2.99 | 2.99 | 1.94 | 1.94 |
| Household goods | 10.05 | 12.22 | 15.39 | 17.55 | 24.78 | 25.67 |
| Household services | 7.66 | 10.41 | 8.84 | 10.77 | 13.64 | 10.84 |
| Childcare | 0.00 | 0.00 | 0.00 | 0.00 | 262.08 | 262.08 |
| Personal goods and services | 17.62 | 31.38 | 18.52 | 38.67 | 30.93 | 40.88 |
| Motoring | 0.00 | 0.00 | 0.00 | 0.00 | 64.33 | 64.86 |
| Other travel costs | 43.30 | 86.65 | 15.42 | 19.16 | 6.30 | 32.49 |
| Social and cultural participation | 44.82 | 75.16 | 48.20 | 81.56 | 52.25 | 63.81 |
| Rent | 95.45 | 106.67 | 78.99 | 86.81 | 86.81 | 86.81 |
| Headline total (excl. rent, childcare) | 231.32 | 381.06 | 209.68 | 322.87 | 330.31 | 432.55 |
| AHC total (excl. childcare, rent, council tax or water) | 206.55 | 350.05 | 184.92 | 290.76 | 298.12 | 393.09 |
| AHC total (excl. fuel costs) | 193.05 | 333.32 | 173.91 | 275.69 | 282.19 | 375.82 |
| Annual AHC total (excl. fuel costs) | 10,039 | 17,333 | 9,043 | 14,336 | 14,674 | 19,543 |

*Appendix: Table 2. Typical household composition in Nottingham, corresponding MIS composition and assumed AHC annual MIS budget (ONS, 2022b)*

| **Household composition** | **Proportion of Nottingham households (%)** | **Corresponding MIS household composition** | **Assumed MIS budget (£/year)** | **Occupancy rank** |
|---|---|---|---|---|
| *One-person household: Aged 66 years and over* | 10.83 | *Single pensioner* | 9,043 | 1 |
| *One-person household: Other* | 22.81 | *Single* | 10,039 | 1 |
| *Single-family household: Lone parent family, all children non-dependant* | 4.49 | *Single* | 10,039 | 1 |
| *Other household types: Other related household, other family composition* | 1.30 | *Single* | 10,039 | 1 |
| *Other household types: Other, including all full-time students and all aged 66 years and over* | 8.40 | *Single* | 10,039 | 1 |
| *Single-family household: Lone parent family with dependant children* | 9.18 | *Lone parent:1 child* | 14,674 | 2 |
| *Other household types: With dependant children* | 3.20 | *Lone parent:1 child* | 14,674 | 2 |
| *Single-family household: All aged 66 years and over* | 4.89 | *Pensioner couple* | 14,336 | 2 |
| *Single-family household: Married or civil partnership couple, no children* | 7.58 | *Couple* | 17,333 | 2 |
| *Single-family household: Married or civil partnership couple, all children non-dependant* | 3.89 | *Couple* | 17,333 | 2 |
| *Single-family household: Cohabiting couple family, no children* | 6.77 | *Couple* | 17,333 | 2 |
| *Single-family household: Cohabiting couple family, all children non-dependant* | 0.77 | *Couple* | 17,333 | 2 |
| *Single-family household: Married or civil partnership couple, dependant children* | 11.47 | *Couple: 1 child* | 19,543 | 3 |
| *Single-family household: Cohabiting couple family, with dependant children* | 4.41 | *Couple: 1 child* | 19,543 | 3 |